\documentclass{article}

     \usepackage[final]{neurips_2019}

\usepackage[utf8]{inputenc} % allow utf-8 input
\usepackage[T1]{fontenc}    % use 8-bit T1 fonts
\usepackage{hyperref}       % hyperlinks
\usepackage{url}            % simple URL typesetting
\usepackage{booktabs}       % professional-quality tables
\usepackage{amsfonts}       % blackboard math symbols
\usepackage{nicefrac}       % compact symbols for 1/2, etc.
\usepackage{microtype}      % microtypography
\usepackage{enumerate}
\usepackage{fancyhdr}
\usepackage{mathrsfs}
\usepackage{bbm}
\usepackage{xcolor}
\usepackage{tikz-cd}
\usepackage{graphicx}
\usepackage{listings}
\usepackage{hyperref}
\usepackage{bm} % package to bold 'math'
\usepackage{hyperref}
\hypersetup{
    colorlinks=true,
    linkcolor=blue,
    filecolor=magenta,      
    urlcolor=cyan,
}
\newtheorem{remark}{Remark}

\newtheorem{theorem}{Theorem}
\newtheorem{proposition}{Proposition}[section]
\newtheorem{corollary}[proposition]{Corollary}
\newtheorem{lemma}[proposition]{Lemma}
\usepackage{amssymb}

\usepackage{natbib}
\usepackage{amsmath}

\title{$\textit{Alice}$'s Adventures in the Markovian World}

\author{%
  Zhanzhan Zhao\\
  School of Aeronautics and Astronautics\\
  Georgia Institute of Technology\\
  Atlanta, GA 30313 \\
  \texttt{zhanzhan@gatech.edu} 
  \And
  Haoran Sun\\
  School of Mathematics\\
  Georgia Institute of Technology\\
  Atlanta, GA 30313 \\
  \texttt{haoransun@gatech.edu} 
 }

\begin{document}

\maketitle
\begin{abstract}

  This paper proposes an algorithm $\textit{Alice}$ having no access to the physics law of the environment, which is actually linear with stochastic noise, and learns to make decisions directly online without a training phase or a stable policy as initial input. Neither estimating the system parameters nor the value functions online, the proposed algorithm generalizes one of the most fundamental online learning algorithms \textit{Follow-the-Leader} into a linear Gauss-Markov process setting, with a regularization term similar to the momentum method in the gradient descent algorithm, and a feasible online constraint inspired by Lyapunov's Second Theorem.
  The proposed algorithm is considered as a mirror optimization to the model predictive control.
  Only knowing the state-action alignment relationship, with the ability to observe every state exactly, 
  a no-regret proof of the algorithm without state noise is given. The analysis of the general linear system with stochastic noise is shown with a sufficient condition for the no-regret proof. The simulations compare the performance of $\textit{Alice}$ with another recent work and verify the great flexibility of $\textit{Alice}$.

\end{abstract}

\section{Introduction}
\textit{To regret deeply is to live afresh.
 - Henry David Thoreau}
 
Different from the optimal control, which knows the system dynamics, hence has the privilege to stand at the beginning of the time and optimize the cumulative loss up to the terminal time as a function of the control action sequence, different from reinforcement learning, which has the training phase to learn the state-action value function, also different from other online control work which require a stable policy as the initial input to the algorithm (\cite{dean2018regret}, \cite{abbasi2019model}), 
the proposed algorithm $\textit{Alice}$ does not know the physics law of the environment, which is actually linear with stochastic noise, and can only learn to make decisions directly online. She neither estimates the system parameters, nor the value functions.

Only knowing the alignment of her actuators and sensors, with the perfect observer of the state, she would like to achieve a sub-linear regret with respect to a quadratic loss given by the environment at each step after she takes an action.
%, and she would like to approach and stay at the equilibrium point of the environment. 

%$\textit{Alice}$ is named in honor of the novel "$\textit{Alice}$'s Adventures in Wonderland".  

%because it is the first algorithm that resembles human life. 
%We name this type of decision making problem online learning control. 

\textbf{Related Work:} \cite{abbasi2011regret} construct a high-probability confidence set around the system parameters based on online least-squares estimation, and derive the regret bound around $\tilde{O}(\sqrt{T})$ for the first time for the linear quadratic control problem.   However its implementation requires solving a non-convex optimization problem to precision $\tilde{O}(T^{-1/2})$, which can be computationally intractable.
\cite{dean2018regret}  proposes the first polynomial-time algorithm for the adaptive LQR problem that provides high probability guarantees of sub-linear regret. However, in proving the regret upper bound, a stable initial policy is assumed to be given as input. %Moreover, the proposed algorithm in this paper converges a lot faster than the algorithm in \cite{dean2018regret} with the same experiment.  

%online RL side (estimate value functions): assumptions, regret bound  

Instead of the interplay between regret minimization and parameter estimation online, model-free approaches for reinforcement learning (RL) is applied online in \cite{abbasi2019model} to solve the linear quadratic control problem with regret upper bound$O(T^{\frac{2}{3}})$ proved. 
Least-squares temporal difference learning \cite{tu2017least} is used to approximate the state-action value functions online. However, this algorithm also requires a stable policy as input. 

\textbf{Major Contributions:}

1) To the best of our knowledge, the proposed algorithm is the first pure online control algorithm that neither uses any information about the system dynamics, nor owns a stable policy as input, and requires neither parameter tuning nor training phase before testing the algorithm.

2) By deriving the close form expression of the control update, the proposed algorithm is proved to be a no-regret algorithm compared with the optimal control knowing the system dynamics when the linear system has no noise.

3) Theoretical analysis is shown for the general linear system with stochastic noise.

4) The algorithm is compared with  \cite{dean2018regret} with the exact same experiment and the better flexibility of the proposed algorithm $\textit{Alice}$ dealing with online adversarial system variations is verified.

\section{Preliminaries}
The environment is given by 
\begin{equation} \label{x}
    {\bm{x}}_{t+1} = A \bm{x}_t + B \bm{u}_t + \bm{w}_{t+1}, 
\end{equation}
where the state $\bm{x}_t \in \mathbb{R}^{n}$, the agent's action $\bm{u}_t \in \mathbb{R}^{m},$ 
the state transition matrix $A \in \mathbb{R}^{n \times n}$, the state-action alignment matrix $B \in \mathbb{R}^{n \times m}$. $\bm{w}_{t+1}$ corresponds to state noise drawn i.i.d. from an unknown Gaussian distribution $\mathcal{N}(0, W),$ where the covariance matrix $W \in \mathbb{R}^{n \times n}$ is diagonal and $W[i,i] \triangleq {\sigma}[i]^2, i = 1,..,n,$ and $\bm{\sigma} \in \mathbb{R}^{n}.$ 
The initial condition $\bm{x}_0 \sim \mathcal{N}(\bar{\bm{x}}_0, P_0)$.
%which is assumed to be diagonal with $P_0[i,i] \triangleq \sigma_{x_0}[i]^2, i = 1,..,n,$ and ${\bm{\sigma}_{x_0}} \in \mathbb{R}^{n}.$ 

The action $\bm{u}_t$ is parameterized as 
\begin{equation}
    \bm{u}_t = K \bm{x}_t = \Phi_{t}\text{vec}(K),\label{defu}
\end{equation}
where the control gain $K \in \mathbb{R}^{m \times n}, \Phi_t \triangleq \bm{x}_{t}^{\rm T} \otimes I_{m},$  $\otimes$ denotes the kronecker product, and $\text{vec}(\cdot)$ is the vector operator.

% \begin{definition}
% A system is said to be controllable if for any initial state $\bm{x}_0$, there exist a finite positive integer $k$ and input sequence $\bm{u}_0, ..., \bm{u}_{k-1}$ such that $\bm{x}_k = 0.$ \label{def1}
% \end{definition}

% \begin{lemma} 
% $[1]$ If system ${\bm{x}}_{t+1} = A \bm{x}_t + B \bm{u}_t$ is controllable in the sense of Definition \ref{def1}, then there exists a complete state (dynamic) feedback controller $\bm{u}_t = K_t\bm{x}_t$, which can assign the closed-loop eigenvalues $\text{eig}(A+BK_t)$ arbitrarily. \label{control}
% \end{lemma}

The loss function $f_t(K_t)$ is defined as
\begin{equation}
    f_t(K_t)= \frac{\eta}{2} \bm{x}_{t}^{\rm T} \bm{x}_{t} + \frac{\beta}{2} \bm{u}_{t-1}^{\rm T} \bm{u}_{t-1}, \label{ft}
\end{equation}
where $t = 1,2,..., T,$ $T$ is the terminal time step, $\eta \in \mathbb{R}^{+}$ is the state weighting parameter, and $\beta \in \mathbb{R}^{+}$ is the control weighting parameter. $\frac{\beta}{\eta} = 0$ means that $\textit{Alice}$ would like to achieve her goal at any cost. The larger $\frac{\beta}{\eta}$ is, the less control effort $\textit{Alice}$ would like to pay to achieve her goal. 
%We specially define that $f_0 = \frac{\eta}{2} \bm{x}_{0}^{\rm T} \bm{x}_{0}.$ 

% \begin{remark}
% Notice that $f_t(K(t))$ is usually defined to be
% \begin{align}
%     f_t(K(t)) \triangleq \frac{\eta}{2} {x}(t)^{\rm T} {x}(t) + \frac{\beta}{2} {u}(t)^{\rm T}{u}(t) \label{fake}
% \end{align}
% in other papers on online linear quadratic control, like \href{https://arxiv.org/pdf/1804.06021.pdf}{[2]}. However, at the terminal time $T,$ there is no such $u(T)$ to move the terminal state $x(T)$ to the next time step. Hence \eqref{fake} will lose its meaning at the terminal time. However, the most important motivation of control is to bring the terminal state to the desired state, hence the terminal cost must be defined. So the author proposes \eqref{ft}.
% Another difference between \eqref{ft} and \eqref{fake} is at $t=0.$ \eqref{ft} has no definition at $t=0,$ since $x(0)$ is already a given state.
% \end{remark}

% At each time step $t$, the agent observes the current state $x(t)$ with the corresponding loss $f_t(K(t))$ revealed, takes an action $u(t)$ and moves to the state $x(t+1)$ at time step $t+1$ according to \eqref{x}. The agent is assumed to know $B$ and not know $A$. The action $u(t)$ is decided by the proposed algorithm $\textit{$\textit{Alice}$}$ introduced in this paper. 

% The cost up to time step $t$ is defined as 
% \begin{align}
% \begin{split}
%     J_{1:t}(K) &=  \sum_{i=1}^{t} \frac{\eta}{2} \hat{\bm{x}}_{i}(K)^{\rm T} \hat{\bm{x}}_{i}(K) + \frac{\beta}{2} \hat{\bm{u}}_{i-1}(K)^{\rm T} \hat{\bm{u}}_{i-1}(K).
%      f_t(K_t)
% \end{split}
% \end{align}

The regret up to the terminal step $T$ is defined as
\begin{equation}
     \text{Reg}_{T} = \sum_{t=1}^{T} f_t(K_t) - \sum_{t=1}^{T} f_t(K^{*}),\label{regret}
\end{equation}
where $K^{*} = \lim_{T \to \infty}\arg \min_{K} \frac{1}{T}\sum_{t=1}^{T} \mathbb{E}[f_{t}(K)],$ which is the fixed control gain given by the infinite horizon Linear Quadratic Regulator (LQR).

\textbf{Assumption 1}: It is assumed that the state $\bm{x}_{t}$, $\forall t =0,1,2,...,$ can be observed exactly.

\textbf{Assumption 2}: There exists $K$, such that $||A+BK||_2$ can be placed arbitrarily. (See \cite{abbasi2014tracking} and \cite{ibrahimi2012efficient} for the similar assumptions.)

\textbf{Assumption 3}: $\|A\|_2 < \frac{\eta +\beta}{\beta},$ which intuitively means that there is enough control effort.

%
%
% In this paper $\pi$ is given by the infinite horizon Linear Quadratic Regulator (LQR), where $K^{*} = \lim_{T \to \infty}\arg \min_{K} \frac{1}{T}\sum_{t=1}^{T} \mathbb{E}[f_{t}(\bm{x}_{t}, \bm{u}_{t-1})],$ which is the fixed control gain matrix. 

%where $K^{*} = \lim_{T \to \infty}\arg \min_{K} \frac{1}{T}\sum_{t=1}^{T} \mathbb{E}[f_{t}(K)],$ which is the fixed control gain given by the infinite horizon Linear Quadratic Regulator (LQR).
% \begin{remark},
% As $T \to \infty,$
% $$
% K^{*} = \arg \min_{K}\sum_{t=1}^{T} f_{t}(K),
% $$
% where $K^*$ is the fixed control gain given by the infinite horizon Linear Quadratic Regulator (LQR).
%When $T$ is finite, 
%$K^*$ is given by the finite horizon linear quadratic optimal control law (P-LQN).
% \end{remark}

%The goal of this note is to propose an algorithm $\textit{$\textit{Alice}$}$ with sub-linear regret bound. 

\section{$\textit{Alice}$}
The current time in the environment is $t$. $\textit{Alice}$ now immersed in her fantasy world, is thinking about which fantasy action $\hat {\bm{u}}_{i-1}(K_{t+1}) \triangleq K_{t+1} \bm{x}_{i-1}$ she should have taken when she was at state $\bm{x}_{i-1},$ where $i=1,2,...,t.$ The control gain $K_{t+1}$ is a variable to be optimized using \eqref{k} at time step $t$ and will be applied as the control gain at time step $t$ in the environment.

After imaging which fantasy action $\hat{\bm{u}}_{i-1}(K_{t+1})$ she should have taken at state $\bm{x}_{i-1}$, $\textit{Alice}$ imagines that she has landed at a fantasy next state $\hat{\bm{x}}_{i}(K_{t+1})$ by the following operation in her mind
\begin{align}
    \hat{\bm{x}}_{i}(K_{t+1}) \triangleq  (\bm{x}_{i} - B\bm{u}_{i-1}) + B\hat{\bm{u}}_{i-1}(K_{t+1}). \label{xhat}
\end{align}
Since the actual state transition in the environment at time step $i-1$ followed that $\bm{x}_i = A\bm{x}_{i-1} + B \bm{u}_{i-1} + {\bm{w}_{i}},$ plugging which into \eqref{xhat} yields that
\begin{align}
   \hat{\bm{x}}_{i}(K_{t+1}) =  \mathcal{A}\bm{x}_{i-1} + B\hat{\bm{u}}_{i-1}(K_{t+1}) + {\bm{w}_{i}}. \label{truth}
\end{align}

% \mathcal{A}\bm{x}_{i-1} + B\hat{\bm{u}}_{i-1}(K_{t+1}) + {\bm{w}_{i}}

where $\mathcal{A}$ denotes the state transition matrix $A$ but inaccessible to $\textit{Alice}$ and our analysis. 
Later on $\mathcal{A}$ will be used instead of $A$ to get this paper distinguished from model-based algorithms and analyses.  
Note \eqref{xhat} reveals that although $\textit{Alice}$ cannot access $\mathcal{A}$ and ${\bm{w}_{i}}$, she can still make the correct state transition in her mind to the fantasy next state $\hat{\bm{x}}_{i}(K_{t+1})$ since she has access to $\bm{x}_{i}$ and $B\bm{u}_{i-1}$. 
%The second equality in \eqref{xhat} can be get by substituting \eqref{x} into the first equality in \eqref{xhat}.
%Notably, \eqref{opA} denotes the real dynamics, and \eqref{xhat} denotes the way how $\textit{Alice}$ can get $\hat{\xB}_{i}(K)$ when she does not know ${A}$. 

The fantasy loss function at step $i$ follows that
\begin{align}
    \hat{f}_i(K_{t+1}) &= \frac{\eta}{2} \hat{\bm{x}}_{i}(K_{t+1})^{\rm T} \hat{\bm{x}}_{i}(K_{t+1}) + \frac{\beta}{2} [B\hat{\bm{u}}_{i-1}(K_{t+1})]^{\rm T} [B\hat{\bm{u}}_{i-1}(K_{t+1})].\label{fhat}
\end{align}
%Specially at the initial time, it is defined that $\hat{f}_0(K) = {f}_0(K)$.

At time step $t$ in the environment, after having travelled again from state $\bm{x}_{0}$ to $\bm{x}_t$ in her mind, and imagined having taken a different $\hat{\bm{u}}_{i-1}(K_{t+1})$ at each $\bm{x}_{i-1}$ and received a fantasy next state $\hat{\bm{x}}_{i}(K_{t+1}),$ she sums up all the fantasy losses from $i = 1$ to $t$, which yields the following fantasy cost function
\begin{align}
\begin{split}
    J_{1:t}(K) &=  \sum_{i=1}^{t} \frac{\eta}{2} \hat{\bm{x}}_{i}(K)^{\rm T} \hat{\bm{x}}_{i}(K) + \frac{\beta}{2} [B\hat{\bm{u}}_{i-1}(K_{t+1})]^{\rm T} [B\hat{\bm{u}}_{i-1}(K_{t+1})].
\end{split}
\end{align}

Still at time step $t,$  $\textit{Alice}$ updates her real action in the environment based on the following optimization, where $\bm{u}_t$ are random vectors for $t = 0,1,.., t_{w}-1$, and for $t\ge t_w$
\begin{align}
K_{t+1} &= \arg\min_{K} J_{1:t}(K) + \lambda \|BK - BK_t\|_2  \label{k} \\ 
& \text{subject to}\  
\|\hat{\bm{x}}_{t}(K)\|_2 \leq \alpha ||\bm{x}_{t-1}||_2, t \geq t_c \notag\\
\bm{u}_{t} &= K_{t+1}\bm{x}_t,\label{u}
\end{align}
where the hyper-parameters $\alpha \in (\frac{1}{\gamma},1), \lambda \geq 0$, $t_w \geq 1$ is the initial waiting time, $t_c \geq 1$ denotes the waiting time for the hard constraint to function.
$\gamma >1$ is a parameter which gives the limit performance that we would like $\textit{Alice}$ to approach her goal considering the state noise, and $||\bold{x}_{T}||_2 \leq 3\gamma ||\bm{\sigma}||_2$. $\gamma$ has its natural limit and hence $\gamma >1$ at least.

% denotes the limit capacity of $\textit{Alice}$ to approach her goal considering the state noise, which is a parameter p and the stop criterion for $\textit{Alice}$ can be $||\bold{x}_{T}||_2 \leq \gamma ||\bm{\sigma}||_2$. 

\begin{remark}
Disregarding the hard constraint in \eqref{k} and setting $\lambda = 0$, \eqref{k} will be reduced to the online learning algorithm Follow-the-leader (FTL), but is specially formulated in this paper to generalize to a linear Gauss-Markov process setting.

The intuition of $\textit{Alice}$'s hard constraint is drawn from Lyapunov's Second Theorem, which states that the equilibrium is exponentially stable for a linear system if and only if there exists the Lyapunov function (e.g, $\|\bm{x}_t\|_2$) of the system decreases with time. So to get to the equilibrium point, $\textit{Alice}$ would like to make sure that the fantasy current Lyapunov function $\|\hat{\bm{x}}_{t}(K)\|_2$ in her fantasy decreases compared with the previous $||\bm{x}_{t-1}||_2$ in the reality world.

The soft constraint $\lambda \|BK - BK_t\|_2$ encourages $\textit{Alice}$'s control gain $K_{t+1}$ to stay close to $K_t$ as time increasing.
This is similar to the momentum method in the gradient decent algorithm. 
%
%As we assume our system is strong exponentially stabilizable, every state can approach the origin directly. So, the idea of momentum is helpful.
\end{remark}

\noindent\rule{15cm}{0.8pt}\\
$\textit{Alice} (\bm{x}_{0}, B, T, \eta, \beta, \alpha, \lambda, \gamma)$\\
\noindent\rule{15cm}{0.4pt}\\
For $t = 0$ to $t_w-1:$\\
\hspace*{0.5em} $\textit{Alice}$ takes a random action $\bm{u}_{t}$\\
\hspace*{0.5em} $\textit{Alice}$ observes $\bm{x}_{t+1}, {f}_{t+1}$\\
End For\\
For $t = t_w$ to $T:$\\
 \hspace*{0.5em} If $||\bold{x}_{t}||_2 \leq 3\gamma ||\bm{\sigma}||_2$:\\
 \hspace*{1.5em}  Break;\\
 \hspace*{1.5em}$\textit{Alice}$ takes action  $\bm{u}_{t} = 0.$\\
 \hspace*{0.5em} Else:\\
 \hspace*{1.5em}$\textit{Alice}$ chooses $K_{t+1}$ according to \eqref{k}\\
 \hspace*{1.5em}$\textit{Alice}$ takes action  $\bm{u}_{t}$ according to \eqref{u}\\
 \hspace*{0.5em} End If\\
 \hspace*{0.5em}$\textit{$\textit{Alice}$}$ observes $\bm{x}_{t+1}$ and receives ${f}_{t+1}(K_{t+1}), \hat{f}_{t+1}(K_{t+1})$\\
% \hspace*{0.5em} t = t + 1\\
End For\\
\noindent\rule{15cm}{0.4pt}\\

\subsection{\textit{Alice}, Model Predictive Control and Follow-the-leader}
Another optimization considered as the mirror optimization to $\textit{Alice}$ is the model predictive control (MPC). Written in accordance with the terms defined in this paper yields that, at each time step $t$ with $\bm{x}_t$ observed, 
\begin{align}
    &\min_{U \triangleq \{\bm{u}_{t},..., \bm{u}_{t+N-1}\}} J_{t:t+N-1} = \sum_{i=t+1}^{t+N-1} \frac{\eta}{2} \bm{x}_{i}^{\rm T}\bm{x}_{i} + \frac{\beta}{2} \bm{u}_{i-1}^{\rm T}\bm{u}_{i-1},
    \\
&\text{subject to}\  
\bm{x}_{i+1} = A \bm{x}_{i} + B\bm{u}_{i}, \label{destnity}
\\
&\hspace*{4em} \bm{x}_{\min} \leq \bm{x}_{i} \leq \bm{x}_{\max}, i= t+1,...,t+N, \label{xcon}\\
&\hspace*{4em} \bm{u}_{\min} \leq \bm{u}_{i} \leq \bm{u}_{\max}, i = t,...,t+N-1, \label{ucon}
\\
& \text{Control update}: \bm{u}_t = \bm{u}^*_t,
\end{align}
where the hard constraints \eqref{xcon} and \eqref{ucon} are optional, $N$ is the horizon of the optimization.

Unlike \textit{Alice}, who thinks in retrospect minimizing the cumulative fantasy loss in her mind (in other words, regrets for her past), MPC minimizes the cumulative loss of the future $N$ steps (in other words, predicts and plans for the future). But since the nature of the future is its uncertainty, MPC, making a hidden but strong assumption on the model for the future \eqref{destnity}, is very sensitive to model uncertainties. To the contrary, due to the nature of the past, which is its certainty, \textit{Alice} can always get the most "genuine" fantasy states, see \eqref{truth} and \eqref{xhat}.

Another nature of the past is its irreversibility, so \textit{Alice} can only regret in her mind. She hopes to enforce Lyapunov's Second Theorem to get the sequence $\|\bm{x_t}\|_2$exponentially decayed, which also implies a no-regret algorithm, but because she does not have the ability to predict the future,  she can only enforce it in her fantasy: she always thinks in retrospect that she should have taken a different action to make the fantasy current state $\|\hat{\bm{x}}_{t}(K)\|_2$ smaller than the  previous $||\bm{x}_{t-1}||_2$.

Given the ability to predict the future, \textit{Alice}'s mirror optimization MPC has the advantage in terms of planning ahead. For example, MPC can plan to make the state and control actions being bounded as wanted in \eqref{xcon} and \eqref{ucon}, but \textit{Alice} cannot even surely bound the state asymptotically in the following analysis in section 4.3.

Although keeps regretting, \textit{Alice} is not a "loser" because she can use her experience from \eqref{k} to improve her future decision making with \eqref{u}. Such spirit is also described by the online learning algorithm FTL \cite{shalev2012online}, and perhaps, Henry David Thoreau. The regret bound of FTL is $O(\log T)$ in that the loss function $f_t(K_t)$ is assumed to be $L_t$ Lipschitz upper bounded by a constant $L,$ which is equivalent to saying that the state $\bm{x}_t$ is assumed to be bounded in our paper according to the definition \eqref{ft}. However, it is not a valid assumption for a Markov-decision process framework, which can be seen from \eqref{x} that $\bm{x}_t$ can blow up exponentially as time increasing. 
To reconcile such conflict, other recent work on online control thus assume a stable policy as input to enable the estimation of system parameters or state-action value functions online. However, our work has fundamental difference from theirs since we do not make such strong an assumption and the goal of \textit{Alice} is to find a stable policy directly online, which is considered as the input to the algorithms in  \cite{dean2018regret} and \cite{abbasi2019model}.

%, to regret deeply from (8) and But the fact that we can only think in retrospect about the past and then modify the future makes everthing lose the sound guarantee and the theory becomes difficult. 

% Future and past are different. Future never really come. Predictions can be fake. Past contains the most real expression of the system. That is why there is a system constraint in MPC's optimization. But there is no system dynamics constraint in Alice's optimization. Actually Alice takes advantage of past and get the most correct sense about the system. Hence MPC has its drawback, and Alice is more flexible. 

%But is it so bad to know the future? Of course not. 
%

\section{Analysis of \textit{Alice}}
\subsection{The Existence of the Optimal Solution and the Feasibility of the Constraints in \eqref{k}}
The Lagrangian $\mathcal{L}_t$ associated with the optimization \eqref{k} at time step $t$ follows that
\begin{align}
\begin{split}
   &\mathcal{L}_t(K, \nu) 
   = \sum_{i=1}^{t}[ \frac{\eta}{2} \hat{\bm{x}}_{i}(K)^{\rm T} \hat{\bm{x}}_{i}(K) + \frac{\beta}{2} [B\hat{\bm{u}}_{i-1}(K_{t+1})]^{\rm T} [B\hat{\bm{u}}_{i-1}(K_{t+1})] ] \\&+ \lambda \|BK - BK_t\|_2 + \nu (\|\hat{\bm{x}}_t(K)\|_2 - \alpha \|\bm{x}_{t-1}\|_2).\label{Lag}
\end{split}
\end{align}

The dual problem to \eqref{k} follows that
\begin{align}
\begin{split}
\sup_{\nu \geq 0} \inf_K \mathcal{L}_t(K, \nu) = \sup_{\nu \geq 0} \inf_K J_{1:t}(K) + \lambda \|BK - BK_t\|_2 + \nu (\|\hat{\bm{x}}_t(K)\|_2 - \alpha \|\bm{x}_{t-1}\|_2).
\label{dual}
\end{split}
\end{align}

\begin{theorem}
Under Assumption 1 and 2, Slater's condition holds true for the optimization \eqref{k} at any time step $t$ with probability at least $\delta = 0.997,$ which also shows that the hard constraints in \eqref{k} are feasible w.p. at least $\delta$. Hence at any time step with probability at least $\delta$, \eqref{k} and its dual \eqref{dual} have strong duality, and such optimal $\nu^*$ and $K^*$ exist that obey the KKT conditions.  \label{it}
\end{theorem}

\textit{Proof:} It follows from Assumption 2 that there exists $K$ such that $||\mathcal{A}+BK||_2$ can be placed arbitrarily. Hence there exists $K$ such that $||\mathcal{A}+BK||_2 = {\delta} < {\alpha - \frac{1}{\gamma}},$
where $\alpha > \frac{1}{\gamma}$ from the definition around \eqref{k}. If follows from \eqref{xhat} that $||\hat{\bm{x}}_{t}(K)||_2 = ||(\mathcal{A}+BK)\bm{x}_{t-1} + \bm{w}_{t}||_2 \leq ||\mathcal{A}+BK||_2 ||\bm{x}_{t-1}||_2 + ||\bm{w}_{t}||_2 < \alpha ||\bm{x}_{t-1}||_2 - \frac{||\bm{x}_{t-1}||_2}{\gamma} + ||\bm{w}_t||_2.$ Since $||\bm{w}_t||_2 \leq ||3\bm{\sigma}||_2$ w.p at least $\delta,$ and $||\bm{x}_{t-1}||_2 > 3\gamma||\bm{\sigma}||_2,$ we could know that the hard constraint $\|\hat{\bm{x}}_{t}(K)\|_2 \leq \alpha ||\bm{x}_{t-1}||_2, t \geq 1$ can be inactive w.p at least $\delta$ at any time step $t$, which means that Slater's condition holds true. Hence the strong duality and KKT conditions are satisfied w.p at least $\delta$ (\cite{boyd2004convex}).

% \textbf{Fact 1}: 
% \begin{align*}
% \nabla_{\text{vec}K} J_{1:t}(K)  &= \sum_{i=1}^{\rm T} [\Phi_{i-1}^{\rm T} B^{\rm T} B \Phi_{i-1} + \beta \Phi_{i-1}^{\rm T} \Phi_{i-1}] \text{vec}(K) + [\Phi_{i-1}^{\rm T}B^{\rm T}(\bm{x}_{i} - B \bm{u}_{i-1})]
% \end{align*}

% \textbf{Fact 2}: 
% \begin{align*}
% \nabla_{K} J_{1:t}(K)  &= \sum_{i=1}^{\rm T} [B^{\rm T}(A+BK) + \beta K] \bm{x}_{i-1}\bm{x}_{i-1}^{\rm T}
% \end{align*}

% \textbf{Fact 3}:
% \begin{align*}
%     \nabla^2_{\text{vec}K}  J_{1:t}(K) = \sum_{i=1}^{\rm T} [\Phi_{i-1}^{\rm T} B^{\rm T} B \Phi_{i-1} + \beta \Phi_{i-1}^{\rm T} \Phi_{i-1}]
% \end{align*}

\subsection{Special Case: Linear System with No Noise}
\begin{theorem}
Set $t_w = n, t_c = T, \lambda = 0,$ and under Assumption 1, 2 and 3, \textit{Alice} is a no-regret algorithm given $\|\bm{\sigma}\|_2 = 0$.
\end{theorem}

\textit{Proof:}
For this case, $\nabla_{K}\mathcal{L}_t$ at time step $t \geq n$ yields that 
\begin{align*}
    \nabla_{K} \mathcal{L}_t &= \eta \sum_{i=1}^{t}B^{\rm T} [\bm{x}_i - B\bm{u}_{i-1} + BK\bm{x}_{i-1} + \frac{\beta}{\eta}BK\bm{x}_{i-1}]\bm{x}_{i-1}^{\rm T}\\ & = \eta [B^{\rm T}\mathcal{A}+ (1+ \frac{\beta}{\eta})B^{\rm T}BK] \sum_{i=1}^{ t} \bm{x}_{i-1}\bm{x}_{i-1}^{\rm T},
\end{align*}
since $\mathcal{A}\bm{x}_{i-1} = \bm{x}_i - B\bm{u}_{i-1}.$
As stated around \eqref{k}, $\bm{u}_i$ are random vectors for $i = 0,1,.., n-1$, thus with probability 1, $\sum_{i=1}^{t} \bm{x}_{i-1} \bm{x}_{i-1}^{\rm T}$ is invertible, $t\geq n$. It follows from $\nabla_{K} \mathcal{L}_t = 0$ that $B^{\rm T}\mathcal{A}+ (1+ \frac{\beta}{\eta})B^{\rm T}BK= 0.$ The optimal $K_{t+1}$ with the minimal norm follows that 
\begin{align}
    K_{t+1} = -\frac{\eta}{\eta+\beta} (B^{\rm T} B)^{\dagger} B^{\rm T} \mathcal{A},\label{regA}
\end{align}
where $(\cdot)^{\dagger}$ denotes the pseudo inverse. Substituting \eqref{regA} into $\bm{x}_{t+1} = (\mathcal{A} + BK_{t+1}) \bm{x}_{t}$ yields that
%
%
% If $B$ is full column rank, then $\mathcal{A} + BK_{t+1} = \frac{\beta}{\beta + \eta}\mathcal{A}$. Hence $\|\bm{x}_{t+1}\|_2 \leq \frac{\beta}{\eta+\beta}\|\mathcal{A}^k\|_2 \|\bm{x}_t\|_2$. If $\rho(\mathcal{A}) < \frac{\eta +\beta}{\beta}$, $\|\bm{x}_{t+1}\|_2$ will decay exponentially, $\forall t \geq n.$\\
%
%More generally, 
$$\bm{x}_{t+1} = \frac{\eta}{\eta+\beta}(\mathcal{A} - B(B^{\rm T}B)^{\dagger}B^{\rm T} \mathcal{A}) \bm{x}_{t}  + \frac{\beta}{\eta+\beta} \mathcal{A} \bm{x}_{t}$$
Hence
$$\|\bm{x}_{t+1} \|_2 \le (\frac{\eta}{\eta+\beta}\|\mathcal{A} - B(B^{\rm T}B)^{\dagger}B^{\rm T} \mathcal{A}\|_2 + \frac{\beta}{\eta+\beta} \|\mathcal{A}\|_2) \|\bm{x}_{t} \|_2$$
It follows from the regression \cite{montgomery2012introduction} that $\|\mathcal{A} + BX\|_2$ is minimized by $BX = -B(B^{\rm T}B)^{\dagger}B^{\rm T}A$. Thus $\|\mathcal{A}-B(B^{\rm T}B)^{\dagger}B^{\rm T} \mathcal{A}\|_2 \le \|\mathcal{A}+BK^*\|_2$, $ \forall K^*.$ It follows from Assumption 2 and 3 that there exists such $K^*$ that
$$\frac{\eta}{\eta+\beta} \|\mathcal{A}+BK^*\|_2 + \frac{\beta}{\eta+\beta}\|\mathcal{A}\|_2 = \delta < 1,$$
Thus $\|\bm{x}_{t+1} \|_2 \le \delta \|\bm{x}_{t}\|_2$. Hence $\|\bm{x}_{t+1}\|_2$ will decay exponentially, $\forall t \geq n.$
It follows from \eqref{ft} that $f_t(K_t)$ will decay exponentially, hence $\sum_{t=1}^{T} f_t(K_t)$ is bounded by a constant. It follows from the property of LQR \cite{kwakernaak1972linear} that $\sum_{t=1}^{T} f_t(K^{*})$ is also bounded by a constant, thus it follows from \eqref{regret} that $\text{Reg}_{T}$ is bounded by a constant. Thus \textit{Alice} is a no-regret algorithm.

\subsection{General Linear System With Stochastic Noise}
Set $t_w = t_c = 1.$

\begin{lemma} \label{lemmahere}
For a strongly convex function $\mathcal{L}_t(K), ||K^{*} - K|| \leq \frac{2||\nabla \mathcal{L}_t(K)||_2}{\min{||\nabla^2\mathcal{L}_t(K)||_2}},$ where $\nabla \mathcal{L}_t(K^*) = 0.$ \cite{boyd2004convex}
\end{lemma}

\begin{theorem}
At any time step $t \geq 2,$ with probability at least $\delta^2$, $||BK^{\sharp}_{t+1} - BK^{\sharp}_t||_2 \leq \frac{2b ||-2 \nu^{\sharp}_{t} \Phi_{t-2}^{\rm T} B^{\rm T} B \hat{\bm{u}}_{t-2} + \eta \Phi_{t-1}^{\rm T} B^{\rm T} \bm{x}_t + (2 \nu^{\sharp}_{t+1} + \beta) \Phi_{t-1}^{\rm T} B^{\rm T} B \bm{u}_{t-1}||_2}{\|(\eta + \beta)\sum_{i=1}^{\rm T} [\Phi_{i-1}^{\rm T} B^{\rm T} B \Phi_{i-1}] + 2 \nu^{\sharp}_{t+1} \Phi_{t-1}^{\rm T} B^{\rm T} B \Phi_{t-1}\|_2} \triangleq \zeta_t,$ where $\nabla_{\text{vec}K} \mathcal{L}_{t-1}(K^{\sharp}_t, \nu^{\sharp}_t; \lambda \equiv  0) = 0,$ $\nabla_{\text{vec}K} \mathcal{L}_{t}(K^{\sharp}_{t+1}, \nu^{\sharp}_{t+1}; \lambda \equiv  0) = 0,$ and $\|B\|_2 \triangleq b.$ \label{theoremk1}
\end{theorem}

 \textit{Proof:} 
% The gradient of the Lagrangian follows that
% \begin{align}
%     \begin{split}
%     &\nabla_{\text{vec}K} \mathcal{L}_t (K, \nu) = \sum_{i=1}^{\rm T} [\eta \Phi_{i-1}^{\rm T} B^{\rm T} B \Phi_{i-1} + \beta \Phi_{i-1}^{\rm T} \Phi_{i-1}] \text{vec}(K) + [\eta \Phi_{i-1}^{\rm T}B^{\rm T}(\bm{x}_{i} - B \bm{u}_{i-1})] \\
%     & + 2 \nu (\Phi_{t-1}^{\rm T} B^{\rm T} B \Phi_{t-1}) \text{vec}(K).
%     \end{split}
% \end{align}

% The hessian of the Lagrangian follows that
% \begin{align}
%     \begin{split}
%     \nabla^2_{\text{vec}K}  \mathcal{L}_t (K, \nu) =  \sum_{i=1}^{\rm T} [\eta \Phi_{i-1}^{\rm T} B^{\rm T} B \Phi_{i-1} + \beta \Phi_{i-1}^{\rm T} \Phi_{i-1}] + 2 \nu (\Phi_{t-1}^{\rm T} B^{\rm T} B \Phi_{t-1}).
%     \end{split}
% \end{align}

with probability at least $\delta^2$, $\nabla_{\text{vec}K} \mathcal{L}_{t-1}(K^{\sharp}_t, \nu^{\sharp}_t; \lambda \equiv  0) = 0,$ $\nabla_{\text{vec}K} \mathcal{L}_{t}(K^{\sharp}_{t+1}, \nu^{\sharp}_{t+1}; \lambda \equiv  0) = 0$ are true. 
Using  $\nabla_{\text{vec}K} \mathcal{L}_{t-1}(K^{\sharp}_t, \nu^{\sharp}_{t}; \lambda \equiv  0) = 0,  \nabla_{\text{vec}K} \mathcal{L}_{t}(K^{\sharp}_{t}, \nu^{\sharp}_{t+1}; \lambda \equiv  0)$ follows that
\begin{align*}
    &\nabla_{\text{vec}K} \mathcal{L}_{t}(K_t, \nu^{\sharp}_{t+1}) = \sum_{i=1}^{t} [\eta \Phi_{i-1}^{\rm T} B^{\rm T} B \Phi_{i-1} + \beta \Phi_{i-1}^{\rm T} B^{\rm T} B \Phi_{i-1}] \text{vec}(K_t) \\&+ [\eta \Phi_{i-1}^{\rm T}B^{\rm T}(\bm{x}_{i} - B \bm{u}_{i-1})] + 2 \nu^{\sharp}_{t+1} (\Phi_{t-1}^{\rm T} B^{\rm T} B \Phi_{t-1}) \text{vec}(K_t)
    \\
    & = \sum_{i=1}^{t-1} [\eta \Phi_{i-1}^{\rm T} B^{\rm T} B \Phi_{i-1} + \beta \Phi_{i-1}^{\rm T} B^{\rm T} B\Phi_{i-1}] \text{vec}(K_t) + [\eta \Phi_{i-1}^{\rm T}B^{\rm T}(\bm{x}_{i} - B \bm{u}_{i-1})] \\
    & + [\eta \Phi_{t-1}^{\rm T} B^{\rm T} B \Phi_{t-1} + \beta \Phi_{t-1}^{\rm T} B^{T}B \Phi_{t-1}] \text{vec}(K_t) + [\eta \Phi_{t-1}^{\rm T}B^{\rm T}(\bm{x}_{t} - B \bm{u}_{t-1})]
    \\&+ 2 \nu^{\sharp}_{t+1} (\Phi_{t-1}^{\rm T} B^{\rm T} B \Phi_{t-1}) \text{vec}(K_t)
    \\
    & = -2 \nu^{\sharp}_{t} (\Phi_{t-2}^{\rm T} B^{\rm T} B \Phi_{t-2}) \text{vec}(K_{t}) + [\eta \Phi_{t-1}^{\rm T} B^{\rm T} B \Phi_{t-1} + \beta \Phi_{t-1}^{\rm T} B^{\rm T} B \Phi_{t-1}] \text{vec}(K_t) \\&+ [\eta \Phi_{t-1}^{\rm T}B^{\rm T}(\bm{x}_{t} - B \bm{u}_{t-1})] + 2 \nu^{\sharp}_{t+1} (\Phi_{t-1}^{\rm T} B^{\rm T} B \Phi_{t-1}) \text{vec}(K_t)
    \\
    & = -2 \nu^{\sharp}_{t} \Phi_{t-2}^{\rm T} B^{\rm T} B \hat{\bm{u}}_{t-2} + \eta \Phi_{t-1}^{\rm T} B^{\rm T} \bm{x}_t + (2 \nu^{\sharp}_{t+1} + \beta) \Phi_{t-1}^{\rm T} B^{\rm T} B \bm{u}_{t-1}.
\end{align*}
The hessian  $\nabla^2_{\text{vec}K} \mathcal{L}_{t}(K_t, \nu^{\sharp}_{t+1})$ follows that
\begin{align*}
    &\nabla^2_{\text{vec}K}  \mathcal{L}_t (K_t, \nu^{\sharp}_{t+1}) =  (\eta + \beta)\sum_{i=1}^{t} [\Phi_{i-1}^{\rm T} B^{\rm T} B \Phi_{i-1}] + 2 \nu^{\sharp}_{t+1} \Phi_{t-1}^{\rm T} B^{\rm T} B \Phi_{t-1},
    % \\
    % &\|\nabla^2_{\text{vec}K}  \mathcal{L}_t (K_t, \nu^{\sharp}_{t+1})\|_2  = m^2(\eta b^2 + \beta)\sum_{i=1}^{t} [\|\bm{x}_{i-1}\|_2^2] + 2\nu^{\sharp}_{t+1} b^2 m^2\|\bm{x}_{t-1}\|_2^2,
\end{align*}
which is a constant. It follows from Lemma \ref{lemmahere} that $\|B(K_{t+1} -K_t)\|_2 \leq \zeta_t.$

% For unconstraint convex problem, we have
% $$\|x^* - x\|_2 \le \frac{1}{2\|\nabla^2 f(x)\|_2} \|\nabla f(x)\|_2$$
% Now, in our context, $K^\sharp$ is the $x^*$ and $K_t$ is the $x$. So, we have $\|\text{vec}(K^\sharp) - \text{vec}(K_t)\|_2$ is bounded by:
% $$\frac{\| ([(1+2\nu)\Phi_{t-1}^{\rm T} B^{\rm T} B \Phi_{t-1}  + \beta \Phi_{t-1}^{\rm T} \Phi_{t-1}] \text{vec}(K^\sharp) + \Phi_{t-1}^{\rm T}B^{\rm T}(A\bm{x}_{t-1}+\bm{w}_i)) \|_2}{2\|\sum_{i=1}^{\rm T} [\Phi_{i-1}^{\rm T} B^{\rm T} B \Phi_{i-1} + \beta \Phi_{i-1}^{\rm T} \Phi_{i-1}] + 2\nu \Phi_{t-1}B^{\rm T} B \Phi_{t-1}\|_2}$$
% We can see the enumerator is just the last term and the denominator is the accumulation of all the state and controls. Without loss of generosity, we can assume all the states and controls are bounded, then we know the right hand side is going to 0 when t is large enough. More specifically, if $\bm{x}$ and $K$ is uniformly bounded by $M$, and we already know $\|\bm{x}\|_2 \ge \sigma $ when the algorithm is still running, we have:
% $$\text{r.h.s} \le \frac{(m^2 \|B\|_2^2 M^2 + \beta m^2 M^2)M + m\|B\|_2M^2}{(\sigma^2m^2\|B\|_2^2 + \beta m^2 \sigma^2)t} $$

% Also, the 2-norm of vector $K$ is the Frobenius norm of the matrix $K$, which is larger than the spectral norm of matrix $K$. We can get:
% $$\|BK^\sharp-BK_t\|_2 \le \|BK^\sharp - BK_t\|_F = O(\frac{1}{t})$$

\begin{theorem} \label{chi}
At any time step $t \geq 2,$ with probability at least $\delta^2$, $||BK_{t+1} - BK_t||_2 \leq \zeta_t$
\end{theorem}

 \textit{Proof:} 
It follows from Theorem \ref{theoremk1} that $K^{\sharp}_{t+1}$ is the solution to the optimization \eqref{k} with $\lambda \equiv 0,$ whereas $K_{t+1}$ is the solution to the optimization \eqref{k}. Thus 
\begin{align}
    &J_{1:t}(K^\sharp) \le J_{1:t}(K_{t+1}),\\
    & J_{1:t}(K_{t+1}) + \lambda \|BK_{t+1} - BK_t\|_2 \le J_{1:t}(K^\sharp_{t+1}) + \lambda \|BK^{\sharp}_{t+1} - BK^{\sharp}_t\|_2.
\end{align}
Hence
\begin{align*}
    \|BK_{t+1} - BK_t\|_2 &\le \|BK^{\sharp}_{t+1} - BK^{\sharp}_t\|_2 - \frac{1}{\lambda}[J_{1:t}(K_{t+1}) - J_{1:t}(K^\sharp_{t+1})] \leq \zeta_t.
\end{align*}

\begin{theorem} \label{coo}
At any time step $t \geq 2,$ with probability at least $\delta^2$, $\|\bm{x}_{t+1}\|_2 \leq (\alpha + \zeta_t) \|\bm{x}_{t}\|_2.$ 
\end{theorem}

% \textit{Proof:} 
% \begin{align*}
%     \|\bm{x}_{t+1}\|_2
%     &= \|(A+BK_{t+1})\bm{x}_t + \bm{w}_{t+1}\|_2 \\
%     &\le \|(A+BK_{t+2})\bm{x}_t\|_2 + \|B(K_{t+1}-K_{t+2})\bm{x}_t\|_2 + \|\bm{w}_{t+1}\|_2 \\
%     &\le \|\hat{\bm{x}}_{t+1}\|_2 + \|B(K_{t+1}-K_{t+2})\|_2 \|\bm{x}_t\|_2 + \|\bm{w}_{t+1}\|_2 \\
    % &\le (\alpha + O(\frac{1}{t})) \|\bm{x}_t\| + \|\bm{w}_{t+1}\|_2
% \end{align*}
% It follows from the hard constraint in \eqref{k} and Theorem \ref{chi} that with probability at least $\delta^2$
% \begin{align*}
%     \|\bm{x}_{t+1}\|_2 \leq \alpha \|\bm{x}_{t}\|_2 - \|3\sigma\|_2 + \lambda_t \|\bm{x}_{t}\|_2 + \|3\sigma\|_2 = (\alpha + \lambda_t) \|\bm{x}_{t}\|_2.
% \end{align*}

\textit{Proof:} 
It follows from \eqref{xhat} that $\|\hat{\bm{x}}_t\|_2 \geq \|{\bm{x}}_t\|_2 - \|(BK_{t+1} - BK_t)\|_2\| {\bm{x}}_{t-1}\|_2.$ It follows from Theorem \ref{it} and \ref{chi} that w.p at least $\delta^2$, $\alpha \|\bm{x}_{t-1}\|_2 \geq \|\hat{\bm{x}}_t\|_2,$ and $||BK_{t+1} - BK_t||_2 \leq \zeta_t.$ Hence w.p. at least $\delta^2, \|{\bm{x}}_t\|_2 \leq (\alpha + \zeta_t)\|{\bm{x}}_{t-1}\|_2.$

\begin{corollary}
If there exists such $t_s$ that $\zeta_t < 1, \forall t > t_s,$ then  it follows from Theorem \ref{coo} that $\alpha + O(\frac{1}{t}) < 1$, then $\|{\bm{x}}_t\|_2$ will exponentially decay, thus the algorithm \textit{Alice} will be a no-regret algorithm.
\end{corollary}

\section{Experiments}
Since under Assumption 1, 2, and 3, $\textit{Alice}$ as a pure online controller, has no need to tune the algorithm parameters, and requires no special information as initial input, the algorithm parameter settings are uniformly set as: $\gamma = 1.2, \alpha = 0.9, \lambda = 0.001, t_w = 1, t_c = 1.$

\subsection{Experiment 1}
To make good comparisons with previous papers, this paper consider the exact the same example as \cite{dean2018regret}, in which Figure 1 compares the median regret of the robust adaptive method(Robust) over 500 experiments with methods like Thompson Sampling (TS, \cite{abeille2017thompson}), and so on. Other works also use the same example: \cite{abbasi2019model}, \cite{dean2017sample}, and \cite{tu2017least}.

The comparison experiment is carried out on the following problem: the initial state is $\bm{x}_0 = 0,$ and 
\begin{align*}
    A = \begin{bmatrix}
1.01 &  0.01& 0\\ 
0.01 &1.01  &0.01 \\ 
 0&0.01  &1.01 
\end{bmatrix}, B = I_3, \eta = 10, \beta = 1, \bm{\sigma} = \begin{bmatrix}
1\\ 
1\\ 
1
\end{bmatrix},
\end{align*}
which corresponds to a marginally unstable Laplacian system where adjacent nodes are weakly connected.

\begin{figure}[!htb]
\minipage{0.5\textwidth}
  \includegraphics[width=\linewidth]{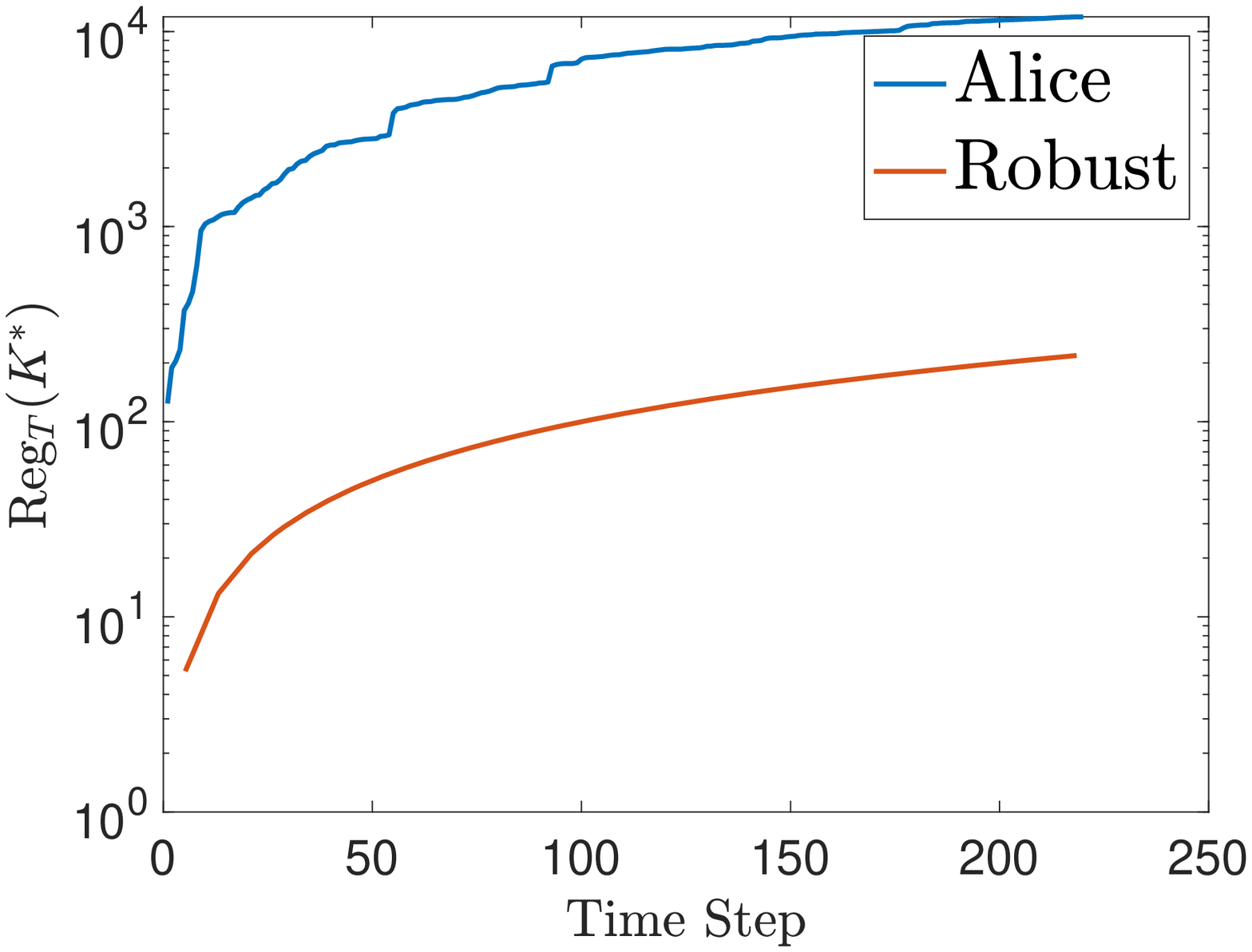}
  \caption{The  median regret comparison.}\label{iloveAlice}
\endminipage\hfill
\minipage{0.5\textwidth}%
  \includegraphics[width=\linewidth]{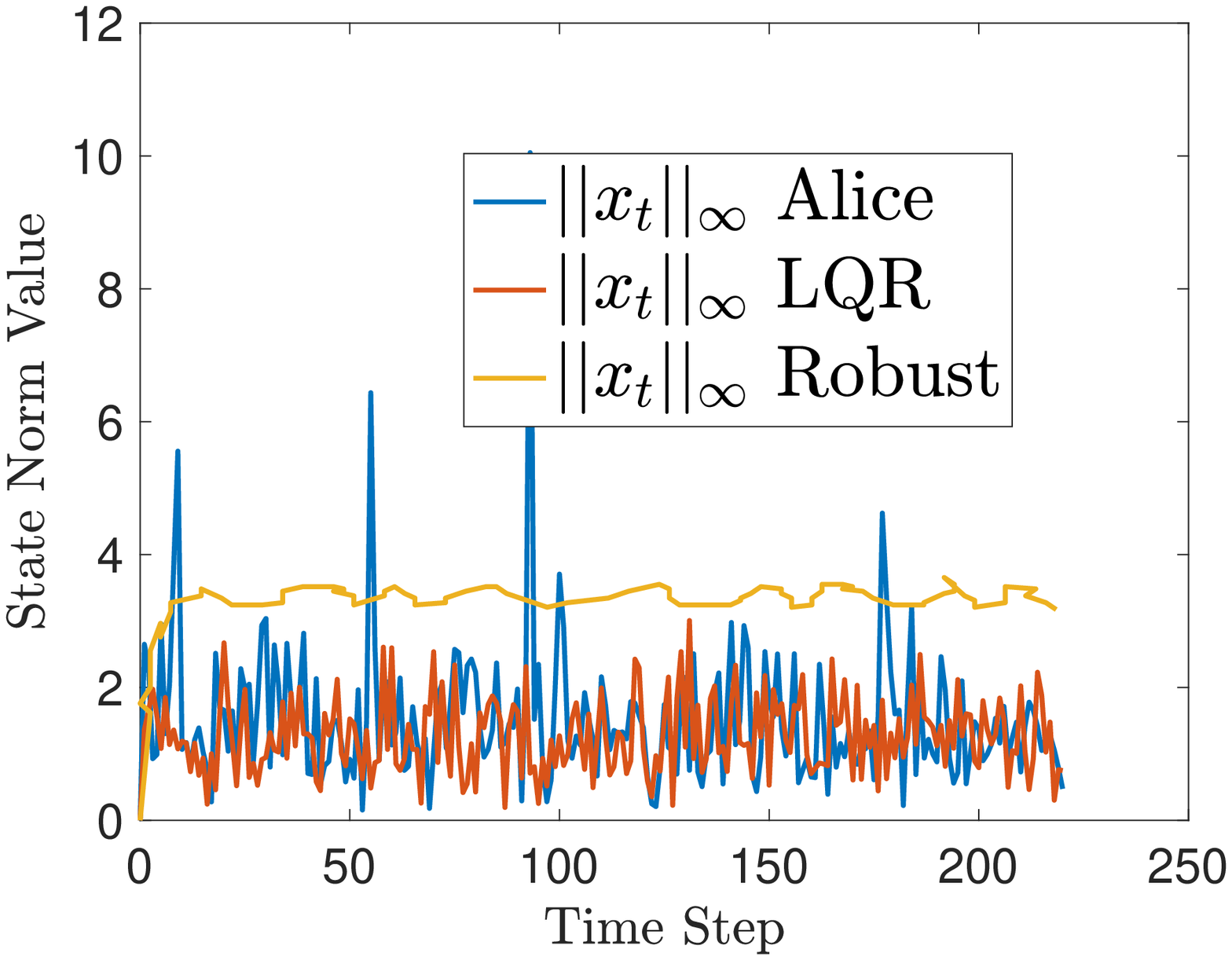}
  \caption{The median state norm comparison.}\label{loveAlice}
\endminipage
% \\
% \minipage{0.5\textwidth}%
%   \includegraphics[width=\linewidth]{regret_noise.eps}
%   \caption{The Regret Propagation}\label{regret_noise}
% \endminipage\hfill
% \minipage{0.5\textwidth}
%   \includegraphics[width=\linewidth]{state_noise.eps}
%   \caption{The State Norm Propagation}\label{Alicemylove}
% \endminipage
\end{figure}

Figure \ref{iloveAlice} shows the median regret comparison between \textit{Alice} and the proposed algorithm in \cite{dean2018regret}, and Figure \ref{loveAlice} shows the state infinity norm comparisons among \textit{Alice}, the proposed algorithm in \cite{dean2018regret} and LQR.

However, it is actually an "unfair" comparison, because Robust has a stabilizing controller as an initial input to the system, and the algorithm adapts the controller on this basis, and $LQR$ knows the system dynamics $A,$ hence has the privilege to predict the future losses. But \textit{Alice} only knows her state-action alignment matrix $B$ without any parameter tuning or training phases. It is the first pure online control algorithm that neither uses any information about the system dynamics, nor owns a stable policy as input. 

Thus, as can be seen in Figure \ref{iloveAlice}, \textit{Alice} suffers great regret in the initial steps compared with Robust. As can be seen in Figure \ref{loveAlice}, \textit{Alice} oscillates more compared with Robust. But the state norm is smaller than Robust on average since the parameter $\gamma$ is set to be 1.2, which means that we would like the steady state norm to be $\|\bm{x}_T\|_2 \leq 3.6\|\bm{\sigma}\|_2 = 3.6.$

\subsection{Experiment 2}
To get distinguished from the online controllers that requires a stable policy as input, we change A matrix in the experiment 1 adversarially online after $t = 10$ to $A^{'}$. To get a more clear sense of the convergence of the algorithm, we reduce the state noise covariance, and change the initial state far from the equilibrium as below. The convergence of the state norm can be seen in Figure \ref{ep2}, the red line as a comparison is also get by $\textit{Alice}$ but without changing the environment dynamics $A$ matrix online. 
Compared with our result, the online controllers using stable policies as initial input are vulnerable to such a case.

\begin{align*}
    A^{'} = \begin{bmatrix}
4.01 &  0.01& 0\\ 
0.01 &1.01  &0.01 \\ 
 0&0.01  &1.01 
\end{bmatrix}, B = I_3, \eta = 10, \beta = 1, \bm{\sigma} = \begin{bmatrix}
0.1\\ 
0.1\\ 
0.1
\end{bmatrix}, \bm{x}_0 = \begin{bmatrix}
5\\ 
5\\ 
5
\end{bmatrix}.
\end{align*}

\subsection{Experiment 3}
\begin{figure}[!htb]
\minipage{0.5\textwidth}
  \includegraphics[width=\linewidth]{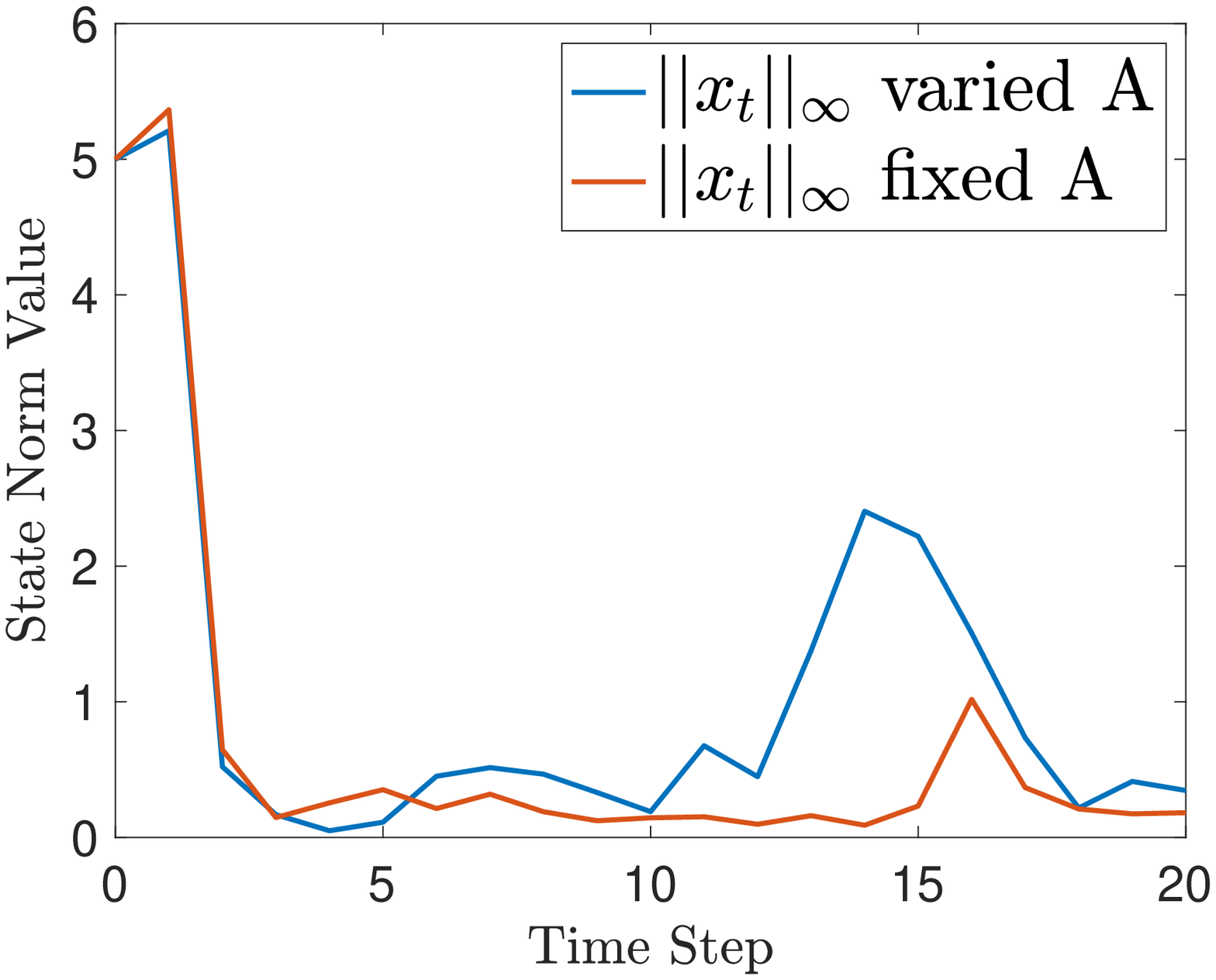}
  \caption{Experiment 2}\label{ep2}
\endminipage\hfill
\minipage{0.5\textwidth}%
  \includegraphics[width=\linewidth]{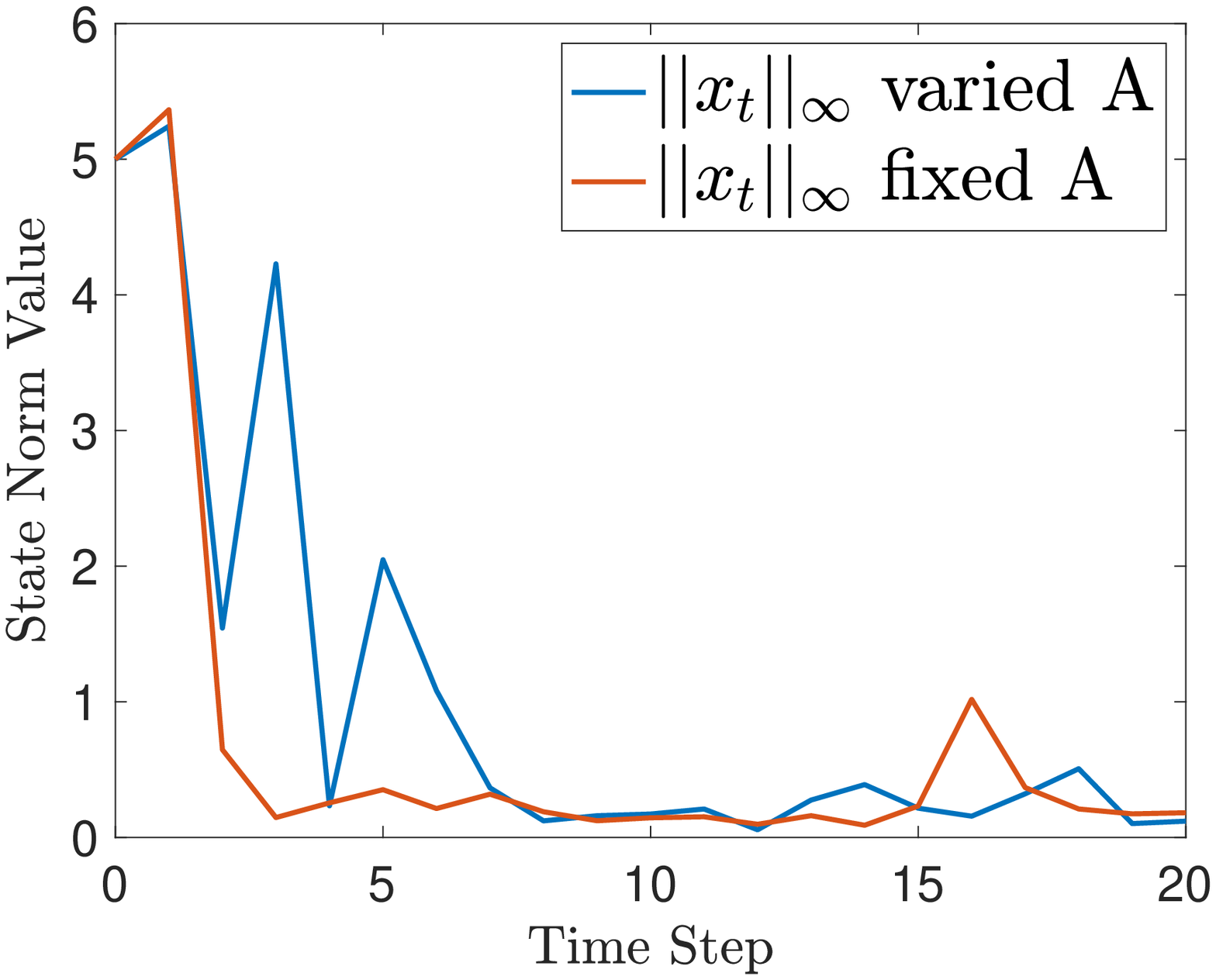}
  \caption{Experiment 3}\label{herehere}
\endminipage
\end{figure}

We further change A matrix in the experiment 1 adversarially online as time varying as below, and the convergence of the state norm can be seen in Figure \ref{herehere}.
\begin{align*}
    A = \begin{bmatrix}
1.01+0.1t &  0.01& 0\\ 
0.01 &1.01  &0.01 \\ 
 0&0.01  &1.01 
\end{bmatrix}, B = I_3, \eta = 10, \beta = 1, \bm{\sigma} = \begin{bmatrix}
0.1\\ 
0.1\\ 
0.1
\end{bmatrix}, \bm{x}_0 = \begin{bmatrix}
5\\ 
5\\ 
5
\end{bmatrix}.
\end{align*}

% \subsection{Experiment 3}
% To further verify the effectiveness of $\textif{Alice}$, we consider the same system as Experiment 1 but only with different initial state and state noise setting: 

\newpage

\bibliographystyle{apalike}
\bibliography{main}

% \begin{thebibliography}{99}

% \bibitem{c1} Liu, Yuan-Ming, and I-Kong Fong. "On the controllability and observability of discrete-time linear time-delay systems." International Journal of Systems Science 43, no. 4 (2012): 610-621.
% \end{thebibliography}

% \subsubsection*{Acknowledgments}

% Use unnumbered third level headings for the acknowledgments. All acknowledgments
% go at the end of the paper. Do not include acknowledgments in the anonymized
% submission, only in the final paper.

\end{document}